\documentclass[iop,apjl]{emulateapj}

\usepackage{apjfonts,graphicx}

\shorttitle{An accretion origin for M31 outer halo globular clusters}
\shortauthors{Mackey et al.}

\begin{document}

\title{Evidence for an accretion origin for the outer halo globular cluster system of M31\altaffilmark{1}}

\author{A.D. Mackey\altaffilmark{2,3}, A.P. Huxor\altaffilmark{4}, A.M.N. Ferguson\altaffilmark{3}, M.J. Irwin\altaffilmark{5}, N.R. Tanvir\altaffilmark{6}, A.W. McConnachie\altaffilmark{7},\\ R.A. Ibata\altaffilmark{8}, S.C. Chapman\altaffilmark{5}, and G.F. Lewis\altaffilmark{9}}

\altaffiltext{1}{Based on observations obtained with MegaPrime/MegaCam, 
a joint project of CFHT and CEA/DAPNIA, at the Canada-France-Hawaii Telescope (CFHT) 
which is operated by the National Research Council (NRC) of Canada, the Institut National 
des Science de l'Univers of the Centre National de la Recherche Scientifique (CNRS) of 
France, and the University of Hawaii.}
\altaffiltext{2}{Research School of Astronomy \& Astrophysics, Australian National 
University, Mt Stromlo Observatory, Weston Creek, ACT 2611, Australia}
\altaffiltext{3}{Institute for Astronomy, University of Edinburgh, Royal Observatory, 
Blackford Hill, Edinburgh, EH9 3HJ, UK}
\altaffiltext{4}{Department of Physics, University of Bristol, Tyndall Avenue, 
Bristol, BS8 1TL, UK}
\altaffiltext{5}{Institute of Astronomy, University of Cambridge, Madingley Road, 
Cambridge, CB3 0HA, UK}
\altaffiltext{6}{Department of Physics \& Astronomy, University of Leicester, 
Leicester, LE1 7RH, UK}
\altaffiltext{7}{NRC Herzberg Institute for Astrophysics, 5071 West Saanich Road, 
Victoria, BC, Canada V9E 2E7}
\altaffiltext{8}{Observatoire de Strasbourg, 11 rue de l'Universit\'{e}, F-67000, 
Strasbourg, France}
\altaffiltext{9}{School of Physics, A29, University of Sydney, NSW 2006, Australia}

\begin{abstract}
We use a sample of newly-discovered globular clusters from the Pan-Andromeda 
Archaeological Survey (PAndAS) in combination with previously-catalogued objects
to map the spatial distribution of globular clusters in the M31 halo. At projected 
radii beyond $\approx30$ kpc, where large coherent stellar streams are readily 
distinguished in the field, there is a striking correlation between these features and
the positions of the globular clusters. Adopting a simple Monte Carlo approach,
we test the significance of this association by computing the 
probability that it could be due to the chance alignment of globular clusters
smoothly distributed in the M31 halo. We find the likelihood of this possibility is 
low, below $1\%$, and conclude that the observed spatial coincidence between
globular clusters and multiple tidal debris streams in the outer halo of M31 reflects 
a genuine physical association. Our results imply that the majority of the remote 
globular cluster system of M31 has been assembled as a consequence of the
accretion of cluster-bearing satellite galaxies. This constitutes the most direct
evidence to date that the outer halo globular cluster populations in some 
galaxies are largely accreted. 
\end{abstract}


\section{Introduction}
It has long been suspected that a significant fraction of the Milky Way's
globular clusters formed in smaller `proto-galactic fragments' that were subsequently 
accreted into the Galactic potential well. First proposed in the seminal
paper by \citet{searle:78}, there has since been gradual accumulation of indirect evidence 
in support of this hypothesis -- modern data suggest that the abundances, ages, velocities, 
horizontal-branch morphologies and sizes of many globular clusters 
at Galactocentric radii $\ga10$ kpc are consistent with an external origin 
\citep[e.g.,][]{zinn:93,mackey:04,marinfranch:09}.
Even so, it has proven problematic to unambiguously identify indvidual clusters
as having been accreted into the Galaxy.
The only {\it direct} observation of this process is the disrupting Sagittarius
dwarf \citep{ibata:94} which is depositing at least five globular
clusters into the Milky Way halo \citep{dacosta:95,martinezdelgado:02,bellazzini:03};
more controversially, the putative Canis Major dwarf may also be responsible
for several new arrivals \citep{martin:04}. 
The extent to which the observed properties of sub-groups within the Milky Way
globular cluster system reflect the assembly history of the Galactic halo thus remains a
critical unresolved question.

As the nearest large spiral galaxy, M31 is an attractive alternative target for studying 
this problem. Its globular cluster system is several times larger than that of the Milky Way, 
and observational investigation of its halo is less prone to the vagaries of projection 
and extinction that plague Galactic surveys. 
It is known that globular clusters projected near its central regions exhibit 
some evidence for sub-clustering in position-velocity space 
that may signal an accretion origin \citep{ashman:93,perrett:03}, although 
interpretation is difficult because of the complex nature of the inner M31 system. 
Potentially less confusing are halo regions at projected radii $R_{{\rm p}}\ga15$ kpc,
where dynamical times are also longer; however, it is only relatively 
recently that these remote areas have been targeted by deep 
wide-field surveys. Various such studies have shown the M31 halo to be littered 
with coherent tidal debris features indicative of one or more accretion events 
\citep{ferguson:02,ibata:07}, and have
also facilitated the discovery of significant samples of remote M31 globular clusters 
\citep{huxor:08} so that it is now possible to begin assessing how these objects 
relate to the stellar halo in this galaxy.

\citet{huxor:10} have derived the first 
radial surface-density profile for M31 globular clusters to extend into the far outer
halo ($R_{{\rm p}}\approx100$ kpc). Their profile exhibits a distinct flattening beyond 
$R_{{\rm p}}\approx30$ kpc, very similar to that observed for the metal-poor field halo, 
and has been interpreted as evidence that accretion processes have played a role in 
building up both components.

In this Letter we use new results from the {\it Pan-Andromeda Archaeological Survey}
\citep[PAndAS;][]{mcconnachie:09} to provide the most extensive map to date of 
globular clusters in the M31 halo, and explore the implications for the origin of this 
system. PAndAS is an ongoing large program on the Canada-France-Hawaii Telescope, 
utilising the MegaCam imager to obtain a deep panoramic view of M31 and M33. 
First-semester imaging and data reduction was completed in mid-2009, revealing in 
exquisite new detail the abundance of low surface brightness substructure present 
in the M31 halo \citep[][see also Fig. 1]{mcconnachie:09}. 

\section{The M31 Globular Cluster Sample}
We consider a globular cluster ensemble defined by confirmed objects in V3.5 of the 
Revised Bologna Catalogue \citep[RBC; e.g.,][]{galleti:07}\footnote[1]{RBC V4.0 was 
released during preparation of the present work. However, the updates 
do not alter our results; in particular, V4.0 does not contain any newly confirmed or
demoted clusters outside $30$ kpc.} 
plus newly-discovered clusters from the first semester of PAndAS imaging.

\begin{figure*}
\begin{center}
\includegraphics[width=168mm]{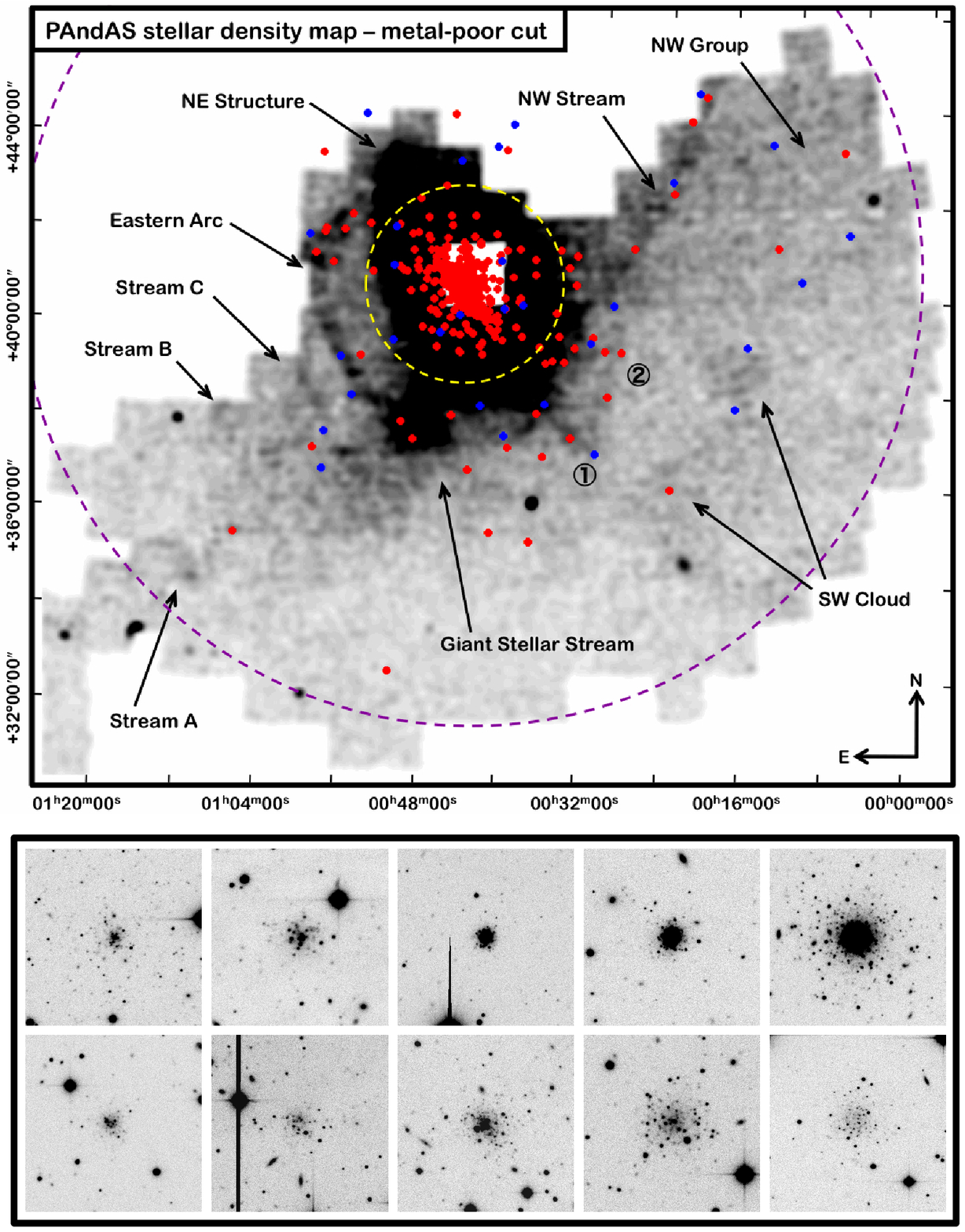}
\end{center}
\caption{First-semester PAndAS map of the spatial density of stellar sources possessing 
luminosities and colours consistent with being metal-poor red-giant branch stars 
($[$Fe$/$H$]\la-1.4$) in the M31 halo \citep[][]{mcconnachie:09}. The two dashed circles,
representing $R_{{\rm p}}=30$ and $130$ kpc, indicate the vast scale of the survey.
Our globular cluster sample is overlaid, marked by red points (compact clusters) and blue 
points (extended clusters). Objects outside the PAndAS area are from our previous survey 
work. Major halo substructres are labelled (see text for details); region (1) indicates the 
ill-defined major-axis feature and nearby overdensities to the east and north, while (2) 
marks the inner western cluster group. The lower panel shows 
$1\arcmin\times1\arcmin$ PAndAS $i$-band thumbnails for ten of our globular clusters 
spanning $30\la R_{{\rm p}}\la120$ kpc and a wide variety of sizes and luminosities.
The lower right-most two are good examples of extended clusters.\label{f:mapthumbs}}
\end{figure*}

In this work we are primarily interested in globular clusters lying outside $R_{{\rm p}}=30$ 
kpc (see below). The RBC list of confirmed globular clusters 
includes $41$ objects discovered in our pre-PAndAS M31 surveys 
\citep{martin:06,huxor:08}, of which $31$ fall beyond $30$ kpc. The RBC also
contains three clusters outside $30$ kpc not discovered by us.
We retain all RBC entries defined as ``extended clusters'' \citep{huxor:05,huxor:08}.
At present there is little evidence that these are anything other than 
{\it bona fide} globular clusters with peculiarly diffuse structures \citep{huxor:10}, 
at least in terms of their constituent stellar populations \citep{mackey:06} and internal 
dynamics \citep{collins:09}.

The extension provided by first-semester PAndAS imaging over the region previously surveyed for 
globular clusters by \citet{huxor:08} consists of nearly complete coverage 
of the inner parts of M31 together with a large halo area to the west and north-west. 
We have searched all fields at $R_{{\rm p}}\ge30$ kpc as well as many fields interior to 
this, using procedures similar to those described by \citet{huxor:08}. This has resulted
in a catalogue of $43$ previously unknown globular clusters, the properties of which 
will be detailed in a forthcoming paper (Huxor et al. 2010b, in prep.). 
For now it is sufficient to note that 
$33$ of these objects fall at projected radii beyond $30$ kpc, including 
$12$ between $50-100$ kpc and four outside $100$ kpc. Sixteen of our 
newly-discovered clusters appear to be of the extended variety.

In summary, our sample contains $67$ clusters with $R_{{\rm p}}\ge30$ kpc, although only $61$ 
of these lie within the present PAndAS footprint.

All PAndAS imaging is taken during dark sky conditions with seeing better than 
$0.7\arcsec$. Under such circumstances globular clusters in the M31 halo partially
resolve into stars, meaning that identification is straightforward and
unambiguous (see Fig. 1).
Following the analysis of \citet{huxor:10}, we believe our search procedure does 
not lead to significant bias or incompleteness in the overall 
cluster selection function down to $M_{{\rm V}}\approx-5$. Furthermore, 
away from the very innermost M31 fields where crowding is non-negligible we expect
no significant spatial variation in completeness.

\section{Analysis and Results}
Figure 1 shows the positions of all globular clusters in 
our sample overlaid on the PAndAS metal-poor ($[$Fe$/$H$]\la-1.4$) stellar density map. 
In the outer parts of the M31 halo where large tidal debris streams are readily 
distinguished ($R_{{\rm p}}\ga30$ kpc), there is a striking correlation between these 
features and the positions of many globular clusters. Indeed, close inspection of
Fig. 1 reveals very few remote clusters 
that do not project onto some kind of underlying field overdensity, even 
though these substructures clearly occupy only a relatively small fraction 
of the overall survey footprint.

In order to put this result on more quantitative ground we undertake several
calculations aimed at estimating the probability that the apparent association 
between clusters and debris streams could be due to chance alignment.
That is, we aim to compute the level of significance at which
similar substructures exist in both the globular cluster and field star distributions
beyond $R_{{\rm p}}\approx30$ kpc.

\subsection{Mock M31 globular cluster systems}
We base our analysis on a simple Monte Carlo methodology utilizing a set of 
$1.5\times10^5$ random realizations of a smoothly-distributed M31 globular 
cluster system. These represent the null case where globular clusters constitute
a well-mixed, unstructured halo population. Comparing the properties of
these mock systems to those of the real M31 system then allows us to assess 
whether the observed globular clusters are indeed spatially correlated with
underlying field overdensities, or whether the apparent association can be ascribed
to stochastic effects.

\begin{figure}
\begin{center}
\includegraphics[width=86mm]{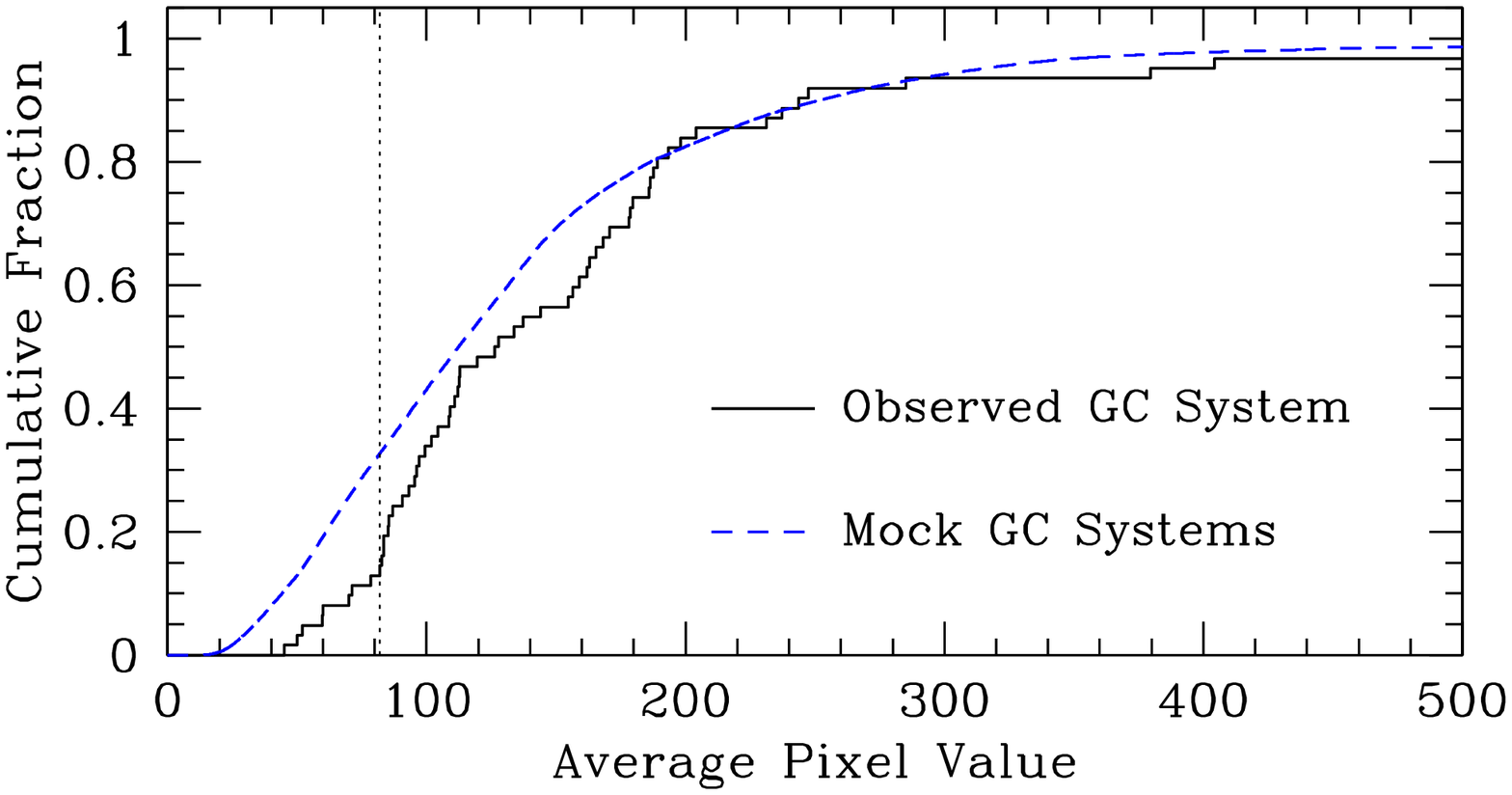}
\end{center}
\caption{Cumulative distributions of mean pixel values about $61$ observed M31 
globular clusters with $R_{{\rm p}}\ge30$ kpc (solid black line)
and clusters in the $1.5\times10^5$ mock systems (dashed blue line). The vertical
dotted line indicates the maximum separation between the two distributions.\label{f:pixeldists}}
\end{figure}
 
In each mock system we randomly generated cluster galactocentric radii 
between $R_{{\rm p}}=30-130$ kpc using a probability distribution function defined 
by the observed globular cluster radial surface-density profile, and selected position 
angles randomly from a uniform distribution such that each individual cluster fell 
within the PAndAS footprint. Our outer limit is the maximum radius with nearly complete 
coverage over the presently-observed area, while our inner limit is the approximate radius 
interior to which there are many overlapping halo features \citep[e.g.,][]{ferguson:02} 
and it becomes meaningless to associate clusters with field substructure via spatial 
coincidence alone. We set the total number of clusters in each mock system equal to the size 
of the known M31 sample between $R_{{\rm p}}=30-130$ kpc within the PAndAS area 
($61$ objects). Our derived cluster surface-density profile closely matches that
obtained by \citet{huxor:10}.

\subsection{Direct comparison with the stellar density map}
We utilized a FITS version of the PAndAS stellar density map to test whether the observed 
M31 globular clusters are preferentially projected against regions with higher densities of 
metal-poor red giant stars. The FITS map has an embedded World Coordinate System,
which allowed us to easily calculate the mean value in a $7\times7$ pixel box about the position 
of each cluster. The clusters themselves are not generally visible on the map since they are 
usually unresolved by the cataloguing software. To be sure, we omitted the central (cluster) 
pixel from the average (one pixel\ $\approx350$ pc). Our box corresponds to
a $\sim2.5\times2.5$ kpc region on the sky, representing an adequate compromise
in obtaining enough pixels without considering an unreasonably broad area about 
each cluster. It is also sufficiently larger than the $\sim2.5$ pixel Gaussian smoothing 
kernel used while generating the map. We considered the mean value in each box, rather 
than all pixels individually, because this smoothing means that adjacent pixel values
are not independent.

We repeated this process for all real and mock globular clusters, and formed the
results into two cumulative distributions (Fig. 2).
These have quite different shapes -- in particular, that for the real cluster system clearly
contains fewer low values than does the distribution for the mock systems.
In other words, the observed M31 globular clusters {\it do} preferentially project onto regions
of higher field star density than would be expected if they constituted an unstructured
halo population. A simple Kolmogorov-Smirnov test provides an estimate of the significance
of this result: the probability that the two distributions were drawn from the same parent
distribution is only $\approx1.3\%$.

This straightforward calculation provides a clear quantitative demonstration that the 
global spatial coincidence between globular clusters and tidal debris streams visible in 
Fig. 1 is almost certainly not due to chance alignment. 

\subsection{Stream-by-stream analysis}
\label{ss:streams} 
Further inspection of Fig. 1 suggests a more complicated picture: some 
streams apparently possess more clusters for their size than do others, while several 
are devoid of clusters altogether. To investigate this in more detail we considered
each major halo overdensity individually. For now, we restrict 
ourselves to only those stellar substructures previously published in the 
literature. These are highlighted in Fig. 1: the 
giant stellar stream \citep{ibata:01}; the northeast structure \citep{zucker:04,ibata:05}; the 
four minor-axis tangent streams A-D \citep{ibata:07}, the latter of which extends to
a coherent arc to the east of M31; and the southwest 
cloud and northwest stream \citep{mcconnachie:09}. \citet{ibata:07} also describe the 
``major-axis diffuse structure'', an ill-defined feature extending to the southwest 
of M31 (region 1 in Fig. 1).
Although there is evidently an overdensity along the major axis here, unlike the other streams
it is not easily identifiable as a single structure;
indeed there are additional faint overdensities extending east and north of this 
feature. We defer analysis of this region until it has been properly characterized,
but note that numerous globular clusters ($\approx8-10$) clearly project onto it.

\begin{figure}
\begin{center}
\includegraphics[width=86mm]{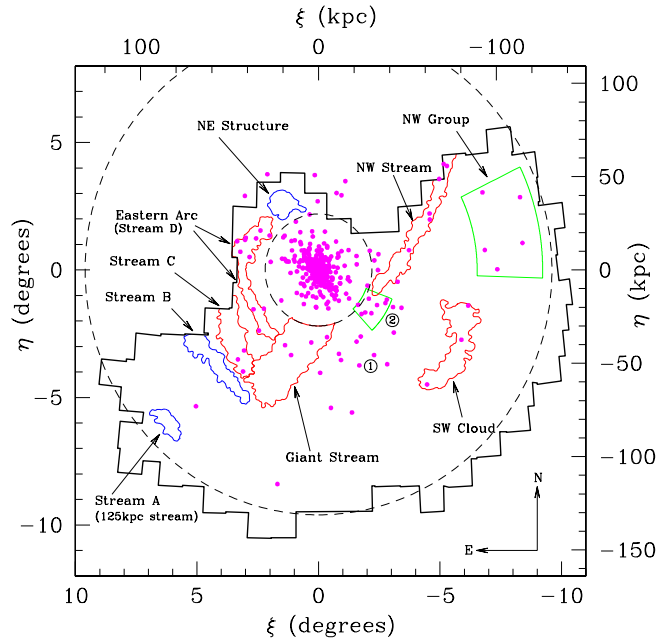}
\end{center}
\caption{Major substructures in the M31 halo. Features associated with multiple/zero 
clusters are outlined in red/blue, while the two cluster overdensities are bounded in green. 
Clusters are magenta points. The dashed circles indicate $R_{{\rm p}}=30$ and $130$ kpc.\label{f:structures}}
\end{figure}
 
We utilized FITS versions of both our metal-poor stellar density map, and an additional map 
including stars with photometric metallicities $-1.4\la[$Fe$/$H$]\la-0.7$, to delineate the
edges of each substructure. These more metal-rich stars alter the appearance of several of the 
overdensities \citep[e.g.,][]{ibata:07,mcconnachie:09}: 
most notably the giant stream, which increases in radial extent and fans westward;
the southwest cloud, which becomes more prominent; and stream C, which is
considerably broadened -- indeed, this feature is known to consist of two distinct
overlapping components \citep{chapman:08}.

\begin{deluxetable*}{lclcccccc}
\tabletypesize{\scriptsize}
\tablecaption{Fraction of mock systems matching the observed cluster-stream associations.\label{t:results}}
\tablehead{
\colhead{Substructure} & \hspace{0mm} & \colhead{Comment} & \hspace{1mm} & \colhead{$N_{{\rm gc}}$} & \hspace{1mm} & \colhead{Number of} & \hspace{0mm} & \colhead{Fraction of}\\
\colhead{} & \hspace{0mm} & \colhead{} & \hspace{1mm} & \colhead{} & \hspace{1mm} & \colhead{matching systems} & \hspace{0mm} & \colhead{matching systems}}
\startdata
\sidehead{{\bf Global system}}
\hspace{2mm} All substructures & & $-$  & & $27+$  & & $372$  & & $0.00248$ \\[1mm]
\hline
\sidehead{{\bf Substructures with multiple clusters}}
\hspace{2mm} Northwest stream  & & excluding boundary cluster    & & $6+$  & & $3\,999$   & & $0.02666$ \\
               & & including boundary cluster    & & $7+$  & & $1\,101$       & & $0.00734$ \\
\hspace{2mm} Southwest cloud   & & $-$  & & $3+$  & & $3\,742$   & & $0.02495$ \\
\hspace{2mm} Eastern arc (Stream D) & & $-$  & & $11+$  & & $791$      & & $0.00527$ \\
\hspace{2mm} Stream C                     & & $-$  & & $3+$  & & $11\,723$  & & $0.07815$ \\
\hspace{2mm} Giant stellar stream    & & $-$  & & $4+$  & & $124\,893$ & & $0.83262$ \\[1mm]
\hline
\sidehead{{\bf Substructures with no clusters}}
\hspace{2mm} Northeast structure  & & $-$  & & $0$  & & $27\,383$   & & $0.18255$ \\
\hspace{2mm} Stream A ($125$ kpc stream) & & $-$   & & $0$  & & $138\,674$  & & $0.92449$ \\
\hspace{2mm} Stream B         & & $-$  & & $0$  & & $88\,828$   & & $0.59219$ \\[1mm]
\hline
\sidehead{{\bf Globular cluster overdensities not associated with identified substructure}}
\hspace{2mm} Western group & & any PA; relaxed bounding box & & $8+$  & & $7\,033$ & & $0.04689$ \\
\hspace{2mm} Northwest group & & any PA; relaxed bounding box & & $5+$  & & $8\,941$ & & $0.05961$ \\[-2mm]
\enddata
\end{deluxetable*}
 
For each substructure we measured the mean and standard deviation of pixels in numerous 
nearby regions, and defined the edge of the structure by following a contour level 
$\approx3.5\sigma$ above the local background. No globular clusters were overplotted 
during this process. Our results are shown in Fig. 3. Although the edges
of these substructures are by nature difficult to define, they are all sufficiently unambiguous
not to alter the conclusions we draw below.

We next counted the fraction of mock systems in which $N_{{\rm gc}}$ or more globular 
clusters overlap spatially with a given substructure, where $N_{{\rm gc}}$ is the number of 
clusters observed to project onto that feature in the real M31 halo. We also considered the 
system globally, grouping together all the identified substructures.

Our results are summarized in Table 1. Taking all
substructures together, we find it very unlikely that the observed spatial overlap of $27$ clusters 
with these features can be explained by random alignment: the probability sits at just $\sim0.25\%$.
This strongly reinforces the result of our previous calculation involving the average local stellar densities.

Individually, the northwest stream and eastern arc are particularly well endowed with clusters.
The fraction of mock systems in which $\ge6$ clusters fall within the northwest stream
is $\sim2.7\%$; this falls to below $1\%$ if the additional boundary-straddling cluster to the 
south is included. Similarly, the frequency with which at least the $11$ observed 
clusters fall within the eastern arc is $\sim0.5\%$ in the mock systems.

For the southwest cloud our calculated probability sits at $\sim2.5\%$, while for stream C it
is less significant at $\sim7.8\%$. Notably however, stream C is the only case where 
velocity information exists for both the field substructure and an associated cluster (EC4),
unambiguously linking the two \citep{collins:09}.

The giant stream is notable as the only structure where the observed number of globular 
clusters approximately matches the number expected in a smoothly-distributed system. 
Since this stream is by far the most luminous stellar substructure in the M31 halo, it appears
significantly underabundant in clusters compared with the streams described above. 
This is perhaps not too surprising: globular clusters located in the outskirts
of the giant stream progenitor may well have been stripped away on earlier orbits about M31.
Furthermore, the globular cluster specific frequencies\footnote[2]{Number of clusters per unit $V$-band 
luminosity, normalized at $M_{V}=-15$.} of dwarf galaxies
span a large range $\sim0-30$ \citep[e.g.,][]{miller:07,peng:08}, so the progenitor might
have possessed comparatively few globulars to start with.

What about the three stellar substructures not associated with any globular clusters? 
Although this occurs commonly in the mock systems for all three features, we have already
assembled ample evidence that clusters are not smoothly distributed in the outer parts of M31. 
Therefore, these three substructures simply demonstrate that not all field overdensities are
necessarily associated with clusters. Streams A and B may be remains of low-mass 
satellites, or they may be `shells' due to the impact of a relatively large accreted galaxy 
\citep{fardal:08,mori:08}. Either way, their intrinsic low luminosity likely explains their
paucity of clusters. It is also perhaps unremarkable that no members of the halo cluster
population are associated with the northeast structure, which observations suggest
is a transient feature in the M31 extended disk \citep[e.g.,][]{ibata:05,richardson:08}.

\subsection{Globular cluster overdensities}
\label{ss:clustergroups} 
Finally, we highlight two globular cluster overdensities that are evident in Figs. 1 
and 3 but not obviously coincident 
with underlying substructures. One sits to the west of M31 at $R_{{\rm p}}\approx35$ 
kpc, and the other to the northwest at $R_{{\rm p}}\approx105$ kpc. They represent the
two groupings with the largest ratio of local cluster density to the azimuthal average
at given radius (a factor $\approx10$ enhancement). The grouping with the third-highest
ratio ($\approx 7.5$) overlaps the upper portion of the eastern arc.

We recognize the {\it a posteriori} nature of our identification of these two overdensities, 
and attempt to quantify their significance fairly by searching the mock systems for aggregations 
at similar radii but unconstrained azimuth, and with local enhancement ratios $\ga5$. With 
these relaxed constraints, indicative probabilities sit at $\approx5\%$ for both ensembles.

This is again fully consistent with globular clusters not being smoothly distributed in the 
outer M31 halo, although here we cannot identify any corresponding field substructures. 
We hypothesize that these two cluster groups may trace underlying substructures that fall below 
the PAndAS low surface brightness limit; confirmation of the nature of these cluster overdensities 
will hence require radial velocity measurements. 
It is intriguing that the very massive globular cluster G1, thought to be the stripped core of a former 
nucleated dwarf \citep[e.g.,][]{bekki:04}, is a prominent member of the western group. 

\section{Discussion}
Put together, our results provide strong evidence that globular clusters in the outer 
halo of M31 are anisotropically spatially distributed, and preferentially associated with underlying 
tidal debris features. Of the $61$ clusters in the PAndAS footprint with $R_{{\rm p}}\ga30$ kpc,
at least $27$ lie on the major substructures outlined in Section \ref{ss:streams}; if the complex major-axis
region is also included, this number rises to $\approx37$. A further $13$ objects are members of 
the two cluster overdensities described in Section \ref{ss:clustergroups}, leaving just $11$ of the 
sample unaccounted for. Figure 1 reveals only a handful of these lie 
away from field overdensities altogether.

These numbers imply that the majority ($\ga80\%$) of the outer globular cluster system of M31 
has been built up via the accretion of satellite host galaxies. This fraction matches closely
that inferred for the outer Galactic system \citep[e.g.,][]{mackey:04,forbes:10}. Our work provides
a striking direct illustration of the \citet{searle:78} paradigm; along with the disrupting Sagittarius 
dwarf, it represents the most clear-cut observation to date of the assembly of a globular cluster 
system in action, and does so on a grand scale -- across an entire galactic system and a huge 
population of clusters.

We note that all but one of the extended clusters in our sample with 
$R_{{\rm p}}\ga30$ kpc are projected onto stellar substructure or are members of a cluster 
overdensity. This observation is consistent with the idea that these puzzling objects may 
predominantly originate in lower-mass galaxies \citep[e.g.,][]{dacosta:09}.

Intriguingly, streams without clusters are seemingly in the minority in
the outer M31 halo. With the caveat that there may be very low-luminosity features falling
below the PAndAS surface-brightness limit, this could imply that most substructure in the 
outer M31 halo is due to the accretion of just a few larger cluster-bearing satellites. 
This would be in qualitative agreement with the results of cosmologically-motivated 
simulations, which suggest that the haloes of large spiral galaxies are predominantly assembled 
via the accretion of a handful ($\la5$) of significant progenitors \citep[e.g.,][]{bullock:05,cooper:10}. 

Owing to their recent discovery, little is yet known about most of the remote M31 clusters 
described in this paper. However, a few have published colour-magnitude diagrams and/or spectra,
which reveal them as almost exclusively metal-poor with $-2.3\la[$Fe$/$H$]\la-1.6$ 
\citep{mackey:06,mackey:07,mackey:10, alvesbrito:09}. 
Others exhibit a small dispersion in integrated colour, further supporting this assertion 
\citep{huxor:10}. This might imply that a significant fraction of metal-poor globular
clusters in large galaxies are accreted objects, as suggested by a variety of models
\citep[e.g.,][]{cote:98,cote:00,prieto:08,muratov:10}. Our present results raise the exciting imminent 
prospect of characterizing, for the first time, the individual properties of multiple accreted 
families of globular clusters in a galactic halo.

\acknowledgements
We are grateful to members of the PAndAS collaboration for commenting on a draft
of this manuscript. ADM acknowledges financial support from the Australian Research Council.
ADM and AMNF acknowledge support by a Marie Curie Excellence Grant from the European 
Commission under contract MCEXT-CT-2005-025869. 

{\it Facilities:} \facility{CFHT (MegaPrime/MegaCam)}.

\end{document}